\documentclass[%
 reprint,
superscriptaddress,
 amsmath,amssymb,
 aps,
]{revtex4-2}

\usepackage{graphicx}
\usepackage{dcolumn}
\usepackage{bm}
\usepackage{color}
\usepackage{multirow}
\usepackage{svg}
\usepackage{arydshln}
\usepackage{float}
\usepackage{soul}

\usepackage{listings} 
\usepackage{booktabs}

\begin{document}

\title{Accurate global machine learning force fields for molecules with hundreds of atoms}

\author{Stefan Chmiela}
\email{stefan@chmiela.com}
\affiliation{Machine Learning Group, Technische Universit\"at Berlin, 10587 Berlin, Germany}
\affiliation{Berlin Institute for the Foundations of Learning and Data -- BIFOLD, Germany}

\author{Valentin Vassilev-Galindo}
\affiliation{Department of Physics and Materials Science, University of Luxembourg, L-1511 Luxembourg City, Luxembourg}

\author{Oliver T.\ Unke}
\affiliation{Google Research, Brain Team, Berlin, Germany}
\affiliation{Machine Learning Group, Technische Universit\"at Berlin, 10587 Berlin, Germany}

\author{Adil Kabylda}
\affiliation{Department of Physics and Materials Science, University of Luxembourg, L-1511 Luxembourg City, Luxembourg}

\author{Huziel E. Sauceda}
\affiliation{Departamento de Materia Condensada, Instituto de F\'isica, 
Universidad Nacional Aut\'onoma de M\'exico, Cd. de M\'exico C.P. 04510, Mexico}
\affiliation{Machine Learning Group, Technische Universit\"at Berlin, 10587 Berlin, Germany}
\affiliation{BASLEARN - TU Berlin/BASF Joint Lab for Machine Learning, Technische Universit\"at Berlin, 10587 Berlin, Germany}
\affiliation{Berlin Institute for the Foundations of Learning and Data -- BIFOLD, Germany}

\author{Alexandre Tkatchenko}
\email{alexandre.tkatchenko@uni.lu}
\affiliation{Department of Physics and Materials Science, University of Luxembourg, L-1511 Luxembourg City, Luxembourg}

\author{Klaus-Robert M\"{u}ller}
\email{klaus-robert.mueller@tu-berlin.de}
\affiliation{Machine Learning Group, Technische Universit\"at Berlin, 10587 Berlin, Germany}
\affiliation{Berlin Institute for the Foundations of Learning and Data -- BIFOLD, Germany}
\affiliation{Google Research, Brain Team, Berlin, Germany}
\affiliation{Max Planck Institute for Informatics, Stuhlsatzenhausweg, 66123 Saarbr{\"u}cken, Germany}
\affiliation{Department of Artificial Intelligence, Korea University, Anam-dong, Seongbuk-gu, Seoul 02841, Korea}

\begin{abstract}

Global machine learning force fields (MLFFs), that have the capacity to capture collective many-atom interactions in molecular systems, currently only scale up to a few dozen atoms due a considerable growth of the model complexity with system size. For larger molecules, locality assumptions are typically introduced, with the consequence that non-local interactions are poorly or not at all described, even if those interactions are contained within the reference \textit{ab initio} data. Here, we approach this challenge and develop an exact iterative parameter-free approach to train global symmetric gradient domain machine learning (sGDML) force fields for systems with up to several hundred atoms, without resorting to any localization of atomic interactions or other potentially uncontrolled approximations. This means that all atomic degrees of freedom remain fully correlated in the global sGDML FF, allowing the accurate description of complex molecules and materials that present phenomena with far-reaching characteristic correlation lengths. We assess the accuracy and efficiency of our MLFFs on a newly developed MD22 benchmark dataset containing molecules from 42 to 370 atoms. The robustness of our approach is demonstrated in nanosecond long path-integral molecular dynamics simulations for the supramolecular complexes in the MD22 dataset.
\end{abstract}

\maketitle


\section{Introduction}

Modern machine learning force fields (MLFFs) bridge the accuracy gap between highly efficient, but exceedingly approximate classical force fields (FF) and prohibitively expensive high-level \emph{ab initio} methods \cite{unke2020,keith2021combining,von2020exploring}. This optimism is based on the universal nature of ML models, which gives them virtually unrestricted descriptive power compared to the statically parametrized interactions in classical mechanistic FFs.
Traditional ML approaches strive towards general assumptions about the problem at hand, such as continuity and differentiability, when constructing models. In principle, any physical property of interest captured in a dataset can be parametrized this way, including collective interactions that are too intricate to extract from the many-body wavefunction. As such, MLFFs can give unprecedented insights into quantum many-body mechanisms~\cite{unke2020}.
Albeit, the exceptional expressive power of global MLFFs goes along with a stark increase in parametric complexity~\cite{zhang2016understanding, allen2019learning, bietti2019inductive} over classical FFs.

As a trade-off, many ML models reintroduce some of the classical mechanistic restrictions on the allowed interactions between atoms. It is unclear to which extent this departure from unbiased ML models compromises their advantages over classical FFs. In particular, localization assumptions are often made to allow a reduction of degrees of freedom across large structures. For example, message-passing neural networks only allow mean-field exchanges between local atomic neighborhoods, which leads to information loss over long distances~\cite{chamberlain2021grand}. This causes a truncation of long-range interactions, which in local models are assumed to have a rather small contribution to the overall dynamics of the system. Nonetheless, it has been shown that long range effects can play a significant role~\cite{ko2021fourth, unke2021spookynet,MBD-science,MBD-scienceadv,unke2022accurate}, limiting the predictive power of local models in nanoscale and mesoscale systems~\cite{chains-NC,chains-PRL}. In fact, several recent MLFF models~\cite{bereau2018non, yao2018tensormol, grisafi2019incorporating, ko2021fourth, unke2019, unke2021spookynet, niblett2021learning, gao2022self} introduce empirical correction terms for specific long-ranged effects (e.g. electrostatics), yet long-range electron correlation effects remain poorly characterized. The number of available MLFF approaches augmented with physical interaction models indicates that we are observing an emerging field which has not yet settled on a universal solution.

\begin{figure}
  \centering
  \includegraphics[width=\columnwidth]{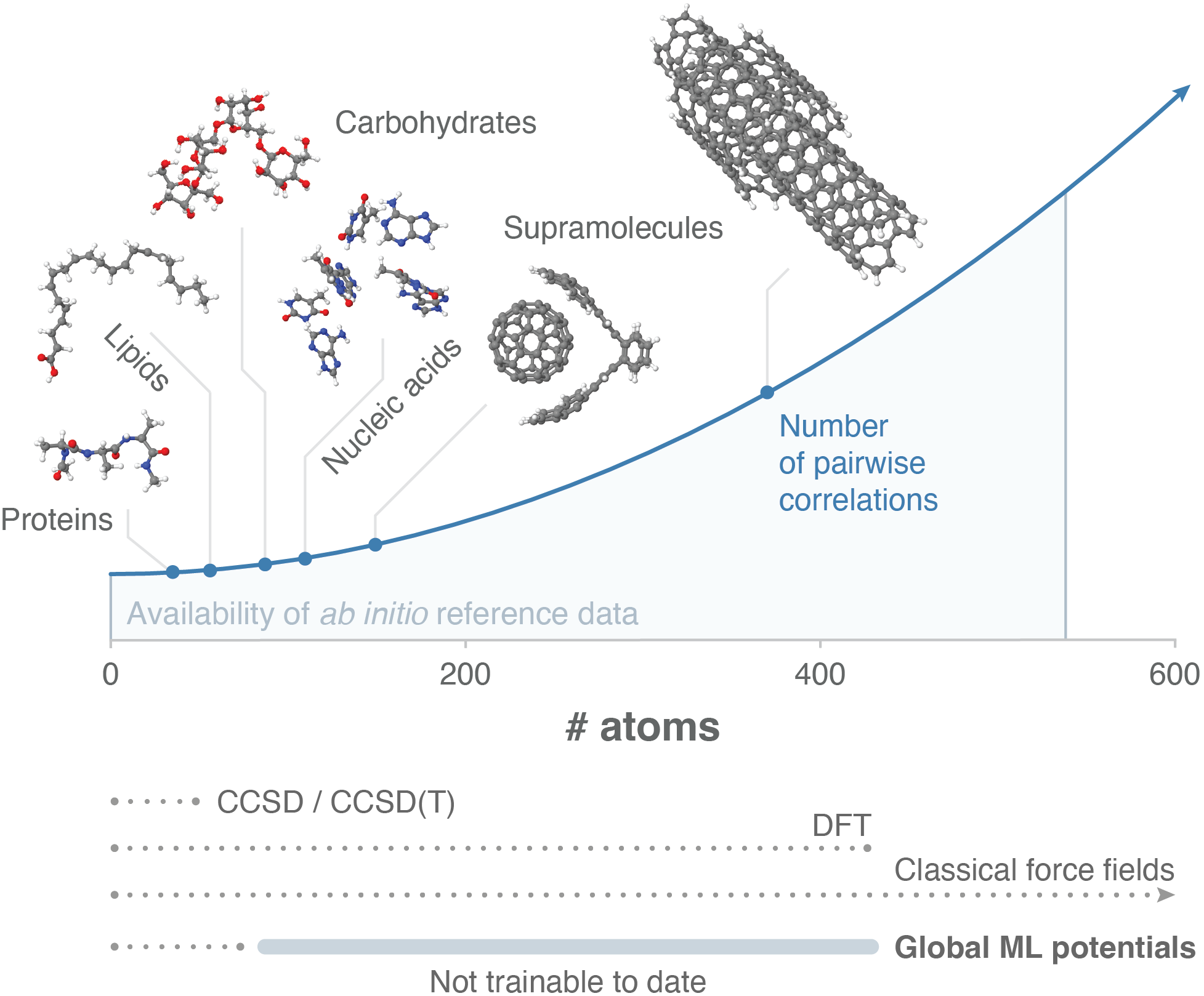}
  \caption{Current global MLFFs only scale to system sizes of a few dozen atoms, restricted by the computational challenge of having to couple a quadratic amount of atom-atom interactions. However, accurate \emph{ab initio} reference data are available for much bigger systems (light blue area). This work scales global models with \emph{ab initio} accuracy to hundreds of atoms, as is demonstrated on examples from four major classes of biomolecules and supramolecules.}
 \label{fig:overview_ml_potential_scales}
\end{figure}

In contrast, global models~\cite{rupp2012fast,chmiela2017,chmiela2018, sauceda2019, sauceda2020molecular, christensen2020fchl} are able to include all interaction scales, but they face the challenge of having to couple at least a quadratic set of atom-atom interactions (see Fig.~\ref{fig:overview_ml_potential_scales}).
Such scaling behavior provides a hard computational constraint and has therefore slowed the development of global models in recent years. Current global models are thus restricted to system sizes of only a few dozen atoms, even though accurate \emph{ab initio} reference data are available for much bigger systems.
Here, we develop a combined closed-form and iterative approach to train global MLFF kernel models for large molecules. Our spectral analysis of these models (see Fig.~\ref{fig:matrix_spectra}) demonstrates that the number of effective degrees of freedom in large molecules is substantially reduced compared to $N^2$ and can be captured using a low dimensional representation~\cite{scholkopf1998nonlinear,braun2008relevant}. Using this insight, our large-scale framework lowers the memory and computational time requirements of the model simultaneously. In a two step procedure, the effective degrees of freedom are solved in closed-form, before iteratively converging the remaining fluctuations to the exact solution for the full problem.

Our focus is on ML models based on Gaussian Processes (GPs), since they posses several unique properties such as linearity and loss-function convexity that can be exploited in pursuit of our goal. We demonstrate the effectiveness of our solution on the symmetric gradient domain machine learning (sGDML) FF~\cite{chmiela2017,chmiela2018}. It allows us to reliably reconstruct sGDML FFs for significantly larger molecules and materials than previously possible~\cite{sauceda2022bigdml}. Our new training scheme can handle systems that contain several hundreds of atoms, all of which are fully coupled within the model. We demonstrate that this parametric flexibility is indeed leveraged to let all atoms participate in generating the energy and force predictions. Our development allows us to study supramolecular complexes, nanostructures, as well as four major classes of biomolecular systems in stable nanosecond-long molecular dynamics (MD) simulations. All of these systems present phenomena with far-reaching characteristic correlation lengths. We offer these datasets as a benchmark (called MD22) that presents new challenges with respect to system size (42 to 370 atoms), flexibility and degree of non-locality. As such, MD22 can be regarded as the next generation of the now well-established MD17 dataset~\cite{chmiela2017}.

\begin{figure*}
  \includegraphics[width=\textwidth]{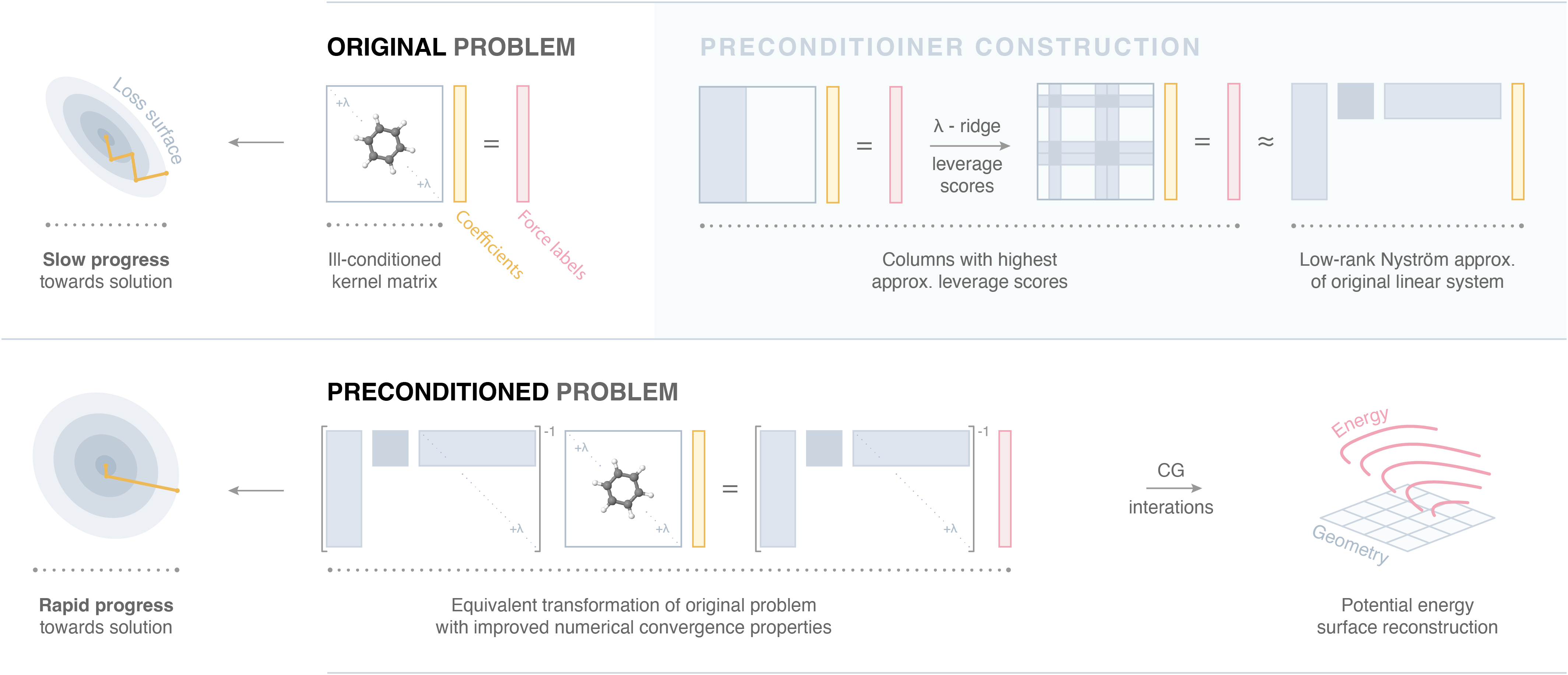}
  \caption{The convergence rate of iterative GP training algorithms is inversely proportional to the condition number of the kernel matrix $\mathbf{K}_\lambda$, which is generally large for strongly correlated many-body systems. Using a preconditioner $\mathbf{P}$, the original linear system is reformulated into an equivalent one $\mathbf{P}^{-1}\mathbf{K}_\lambda \mathbf{\alpha} = \mathbf{P}^{-1}\mathbf{y}$ that is numerically easier to solve. Approximate statistical leverage scores in combination with the Nyström approximation are used to construct a memory efficient and easily invertible representation of $\mathbf{P}$.
}
\label{fig:preconditioning}
\end{figure*}

\section{Large-scale sGDML algorithm}

A data efficient reconstruction of FFs with strong generalization properties requires models that implement the appropriate prior knowledge to compensate for finite reference dataset sizes. Quantum chemical interactions are highly complex and can thus not be fully specified even by massive datasets. We therefore exploit that many complex interactions can be summarized in terms of simple constraints derived from conservation laws  that are far more effective. GPs provide a particularly elegant way of incorporating such constraints, since they are closed with respect to linear transformations such as integral operators or partial differential equations. A combination of linear constraints can be written as a new instance of a GP. This principle is used by the symmetric gradient domain machine learning (sGDML) approach  to construct models that implement all important invariance properties of MLFFs~\cite{chmiela2017}. One of its key characteristics is the use of a kernel $\mathbf{k}\left(\mathbf{x},\mathbf{x}^{\prime}\right) = \nabla_{\mathbf{x}}k_{E}\left(\mathbf{x},\mathbf{x}^{\prime}\right)\nabla_{\mathbf{x}^{\prime}}^{\top}$ that models the force field $\mathbf{f}_{\mathbf{F}}$ as a transformation of an unknown potential energy surface $f_{E}$ such that
\begin{equation}
\boldsymbol{\mathbf{f_{F}}}=-\nabla f_{E}\sim\mathcal{G}\mathcal{P}\left[-\nabla\mu_{E}(\mathbf{x}),\nabla_{\mathbf{x}}k_{E}\left(\mathbf{x},\mathbf{x}^{\prime}\right)\nabla_{\mathbf{x}^{\prime}}^{\top}\right].
\end{equation}
Here, $\mu_{E}: \mathbb{R}^{d} \rightarrow \mathbb{R}$ and $k_{E}: \mathbb{R}^{d} \times \mathbb{R}^{d} \rightarrow \mathbb{R}$ are the respective prior mean and prior covariance functions that define the latent energy-based GP. The training on force examples is motivated by the fact that they are available analytically from electronic structure calculations via the Hellmann–Feynman theorem, with only moderate computational overhead atop energy measurements. Force samples constitute a more efficient way to generate reference data, which sGDML is able to take advantage of~\cite{chmiela2019accurate}. Despite their non-parametric nature, sGDML FFs generally use around one order of magnitude fewer parameters than deep neural network architectures with comparable accuracy, making them computationally less expensive to evaluate (see Appendix~\ref{s:parametric_complexity}). 

\begin{figure}
  \centering
  \includegraphics[width=\columnwidth]{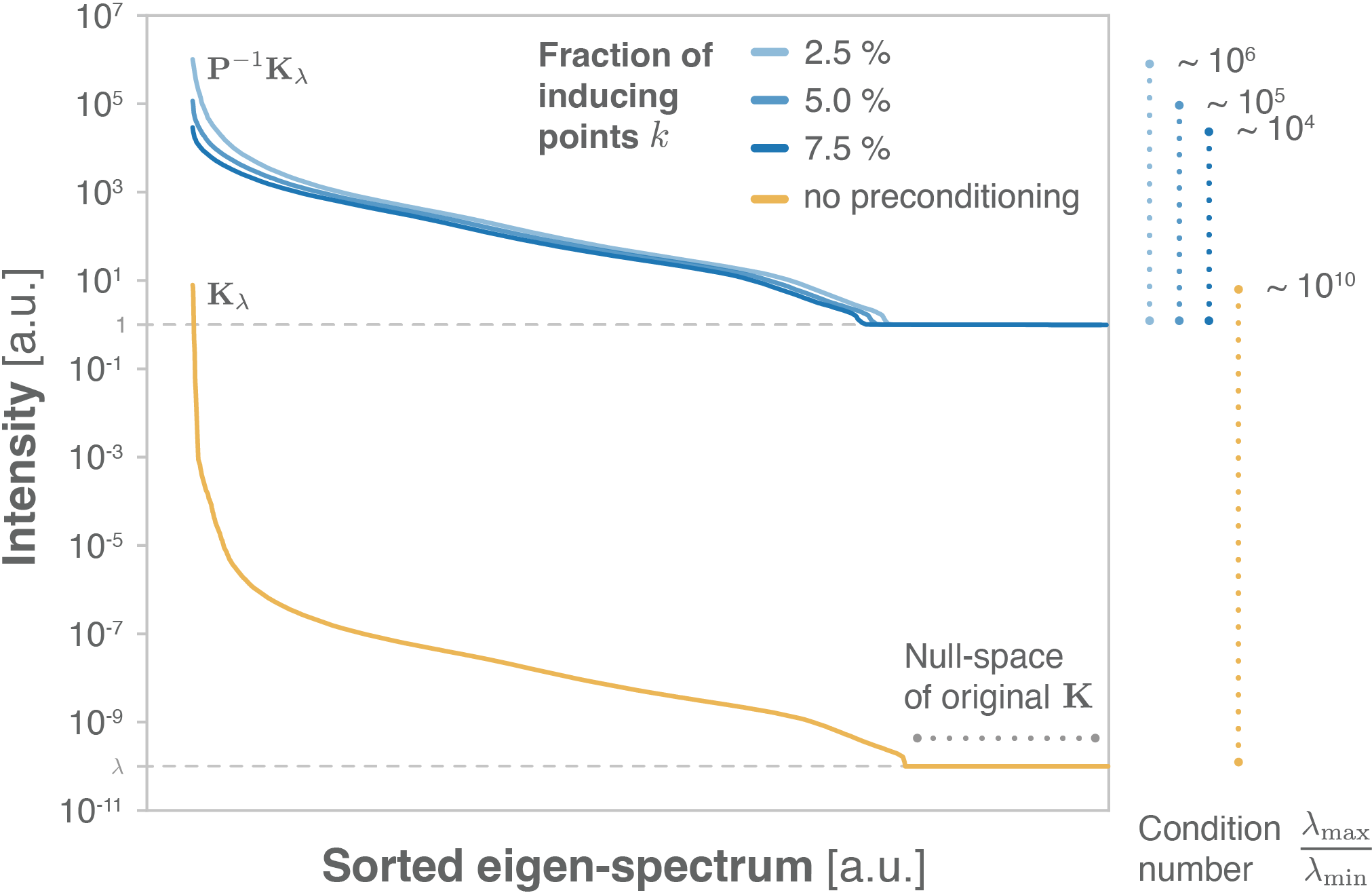}
  \caption{Eigenvalue spectrum of the regularized sGDML kernel matrix $\mathbf{K}_\lambda$ (malonaldehyde, $\sigma = 12$, $\lambda = 10^{-10}$) before and after preconditioning $\mathbf{P}^{-1}\mathbf{K}_\lambda$ with an increasing number of inducing points $k$ (2.5\%, 5\% and 7.5\% of the total number of training points $m$). Note, that regularization lifts the null-space of the original kernel matrix to $\lambda$. Since the same regularization is applied to the preconditioner, the spectrum of $\mathbf{P}^{-1}\mathbf{K}_\lambda$ is thus bounded from below by 1. The dominant part of the spectrum is increasingly attenuated by the preconditioner, reducing the condition number of the matrix and thus accelerating convergence of the CG algorithm.}
 \label{fig:matrix_spectra}
\end{figure}

GPs can be solved in closed form, because their convex loss function yields a linear system when its derivative is set to zero~\cite{williams2006gaussian}. The resulting system $\mathbf{\alpha} = (\mathbf{K} + \lambda \mathbb{I})^{-1}\mathbf{y}$ is symmetric positive-definite by definition of the kernel function, which allows a solution via Cholesky decomposition~\cite{fine2001efficient, bach2002kernel} (see Appendix~\ref{s::cholesky}). While this approach is efficient and numerically stable, it does not scale to large matrix sizes due to its quadratic memory cost.
In search for a solution, a variety of approximation techniques have emerged~\cite{williams2001using, quinonero2005unifying, yang2005efficient, shen2006fast, snelson2006sparse, rahimi2008random, wilson2015kernel, rudi2017falkon}, all drawing on the same insight that kernel matrices often have a comparatively small numerical rank (i.e. a rapidly decaying eigenvalue spectrum; see Fig.~\ref{fig:matrix_spectra}) that can be exploited to formulate a lower-dimensional problem. However, this is only the case with noisy training data, which causes many degrees of freedom of the kernel matrix to become dispensable. \emph{Ab initio} reference calculations are essentially noise-free, giving rise to kernel matrices that can not be compressed without compromising the accuracy of the predictor.~\cite{bach2005predictive}. Therefore, continued efforts are being made towards enabling \emph{exact} large-scale GP inference via numerical gradient-based optimization schemes as opposed to analytical matrix decomposition approaches~\cite{murray2009gaussian,wilson2016deep,gardner2018product,gardner2018gpytorch, wang2019exact}. Those techniques exploit that the kernel matrix $\hat{f}(\mathbf{x})$ only appears as a matrix-vector product in the gradient $\nabla_{\mathbf{\alpha}} \mathcal{L}(\hat{f}(\mathbf{x}), \mathbf{y})$ of the loss function
\begin{equation}
\begin{aligned}
\mathcal{L}\left(\hat{f}(\mathbf{x}), \mathbf{y}\right)=&\left(\hat{f}(\mathbf{x})-\mathbf{y}\right)^{2} + \lambda \|\hat{f}\|^2 \\
=& \; \mathbf{y}^{\top} \mathbf{y} - (2\mathbf{y}^{\top} + \mathbf{\alpha}^{\top} \mathbf{K} + \lambda \mathbf{\alpha}^{\top} )  \mathbf{K}\mathbf{\alpha},
\end{aligned}
\end{equation}
which can be evaluated on-the-fly without ever storing $\mathbf{K}$. Using a gradient-descent scheme, the parameters are then updated until convergence to the exact solution (or as close as necessary) using iterates of the form
\begin{equation}
\mathbf{\alpha}_{t}=\mathbf{\alpha}_{t-1}-\gamma\underbrace{\left[(\mathbf{K} + \lambda \mathbb{I})\mathbf{\alpha}_{t-1} -\mathbf{y}\right]}_{\nabla_{\mathbf{\alpha}} \mathcal{L}(\hat{f}(\mathbf{x}), \mathbf{y})}.
\label{eq:normal_eq_gradient}
\end{equation}
Here, the hyper-parameter $\gamma$ determines the step size (learning rate) with which the minimum of the loss function is approached. In practice, the convergence of this algorithm is severely impeded in regions with large differences in curvature, causing the iterates to jump back and forth across narrow valleys, which slows down progress in some directions on the loss surface (see Fig.~\ref{fig:preconditioning}).

The conjugate gradient (CG) descent algorithm~\cite{hestenes1952methods,saad2003iterative, shewchuk1994introduction} avoids this behavior by exploiting the special topography of GP loss surfaces to make optimal choices of step sizes $\gamma_{t}$ and descent directions $\mathbf{p}_{t}$. It expresses the solution $\mathbf{\alpha}=\sum_{t=1}^{m} \gamma_{t} \mathbf{p}_{t}$ in a basis of $m$ mutually conjugate vectors with respect to $\mathbf{K}_\lambda = \mathbf{K} + \lambda \mathbb{I}$, such that for any pair $\mathbf{p}_i^{\top} \mathbf{K}_\lambda \mathbf{p}_j=0$. Instead of simply following the negative gradient $\mathbf{r}_t = \nabla_{\mathbf{\alpha}} \mathcal{L}(\hat{f}(\mathbf{x}), \mathbf{y})$ as in Eq.~\ref{eq:normal_eq_gradient}, each gradient step is projected away from all previous minimization directions:
\begin{equation}
\mathbf{p}_{t}=\mathbf{r}_{t}-\sum_{i<t} \frac{\mathbf{p}_{i}^{\top} \mathbf{K}_\lambda \mathbf{r}_{t}}{\mathbf{p}_{i}^{\top} \mathbf{K}_\lambda \mathbf{p}_{i}} \mathbf{p}_{i}.
\label{eq:cg_projection}
\end{equation}
Furthermore, rather than fixing the learning rate heuristically, the convexity of the learning problem allows an optimal choice at each iteration step:
\begin{equation}
\gamma_{t}=\frac{\left\langle\mathbf{p}_{t}, \mathbf{r}_{t}\right\rangle}{\left\langle\mathbf{p}_{t}, \mathbf{p}_{t}\right\rangle_{\mathbf{K}_\lambda}}.
\end{equation}

Put together, the iterates $\mathbf{\alpha}_{t+1}=\mathbf{\alpha}_{t}-\gamma_{t} \mathbf{p}_{t}$ look very similar to Eq.~\ref{eq:normal_eq_gradient}. In contrast to basic gradient-descent schemes, this algorithm is parameter-free, foregoing a laborious parameter search. Most importantly, the CG algorithm can be reformulated in a memory efficient way that avoids storing all previous search directions and residual vectors that appear in Eq.~\ref{eq:cg_projection}, using the mutual orthogonality of all $\mathbf{r}_i$ and $\mathbf{p}_i$. For $\mathbf{K}_\lambda$ with fast-decaying eigenvalue spectra (as is almost always the case, see \emph{Marchenko–Pastur distribution}~\cite{marvcenko1967distribution}), the CG
expansion can typically be truncated early (i.e.\ at $t < m$) to reach super-linear convergence in practice~\cite{shewchuk1994introduction}.

One important caveat of the CG algorithm is however its numerical robustness. The number of required iteration steps until convergence $m \sim \mathcal{O}(\kappa)$ is proportional to the condition number $\kappa = \lambda_{\text{max}} / \lambda_{\text{min}}$ (the ratio between largest and smallest eigenvalue) of $\mathbf{K}_\lambda$, which is notoriously bad for correlation matrices with a characteristic steep spectral drop-off. A direct application of this iterative scheme will therefore lead to impracticably slow convergence or even divergence due to numerical errors (see Figs.~\ref{fig:matrix_spectra} and~\ref{fig:convergence_tests})~\cite{wendland_2004, wu2017exploiting}. The standard way of dealing with ill-conditioned problems is to reformulate the original linear system to an equivalent one, for which the iterative solver converges faster. Using a non-singular preconditioning matrix $\mathbf{P}$, we can write
\begin{equation}
\mathbf{\alpha} = ({\mathbf{P}^{-1}\mathbf{K}_\lambda})^{-1}\mathbf{P}^{-1}\mathbf{y},
\end{equation}
to make the convergence depend on the condition number of $\mathbf{P}^{-1}\mathbf{K}_\lambda$ rather than $\mathbf{K}_\lambda$. To continue to satisfy the symmetric and positive definiteness assumptions of the CG algorithm, we require that $\mathbf{P}^{-1} = \mathbf{Q}\mathbf{Q}^{\top}$ exits, although the algorithm can be formulated in a way that does not require an explicit factorization.

In order for any preconditioner to be practical, the cost of solving $\mathbf{P}^{-1}\mathbf{K}_\lambda$ must be low~\cite{gardner2018gpytorch}.
This requirement rules out optimal rank-$k$ approximations obtained from the dominant $k$-dimensional eigenspace of $\mathbf{K}$, as such constructions incur cubic computational cost $\mathcal{O}{(m^3)}$~\cite{cutajar2016preconditioning}. Even algorithms that only compute the leading eigenvalues and eigenvectors are not significantly more efficient, except when $k \ll m$.
Here, we use the Nystr\"om method~\cite{williams2001using} as an alternative, in order to compute a low-cost approximation to that relevant eigenspace. Accordingly, the original kernel matrix is approximated at a cost of only $\mathcal{O}{(k^2 m)}$ from a subset of $k$ of columns as 
\begin{equation}
\mathbf{K} \approx \hat{\mathbf{K}} = \mathbf{K}_{mk} \mathbf{K}_{kk}^{-1} \mathbf{K}_{mk}^{\top}.
\label{eq:nystrom_approx}
\end{equation}
The Woodbury matrix identity~\cite{cutajar2016preconditioning} is then used to efficiently invert $\mathbf{P} = \hat{\mathbf{K}} + \lambda\mathbb{I}$ in parts:
\begin{equation}
\mathbf{P}^{-1} = \lambda^{-1}\left[\mathbb{I}-\mathbf{K}_{mk}\left(\lambda\mathbf{K}_{kk}+\mathbf{K}^{\top}_{mk} \mathbf{K}_{mk}\right)^{-1} \mathbf{K}^{\top}_{mk}\right] .
\label{eq:preconditioner}
\end{equation}
There are many strategies to select the column subspace, the most straightforward of which is simple random sampling~\cite{williams2001using,kumar2012sampling}. Here, we chose a more effective, yet still computationally economical approach based on approximate statistical leverage scores~\cite{golub2013matrix, wahba1990spline, seeger2003fast, foster2009stable, eriksson2018scaling} (see Appendix~\ref{s::app_lev_scores}).

Overall, the accuracy of the approximation $\hat{\mathbf{K}}$ is primarily determined by its dimension $k$ and so is the cost of inverting and applying $\mathbf{P}$, with runtime and memory complexities of $\mathcal{O}(mk^2)$ and $\mathcal{O}(mk)$, respectively. A stronger preconditioner will accelerate convergence, albeit at increased cost per iteration (see Fig.~\ref{fig:convergence_tests}). Note, that the performance and effectiveness of the outlined method crucially depends on a careful numerical implementation, as explained in Appendix~\ref{s::implementation_details}.

\section{Assessment of large-scale molecular force fields}

\paragraph{Convergence properties}

\begin{figure}[b]
  \centering
  \includegraphics[width=1.0\columnwidth]{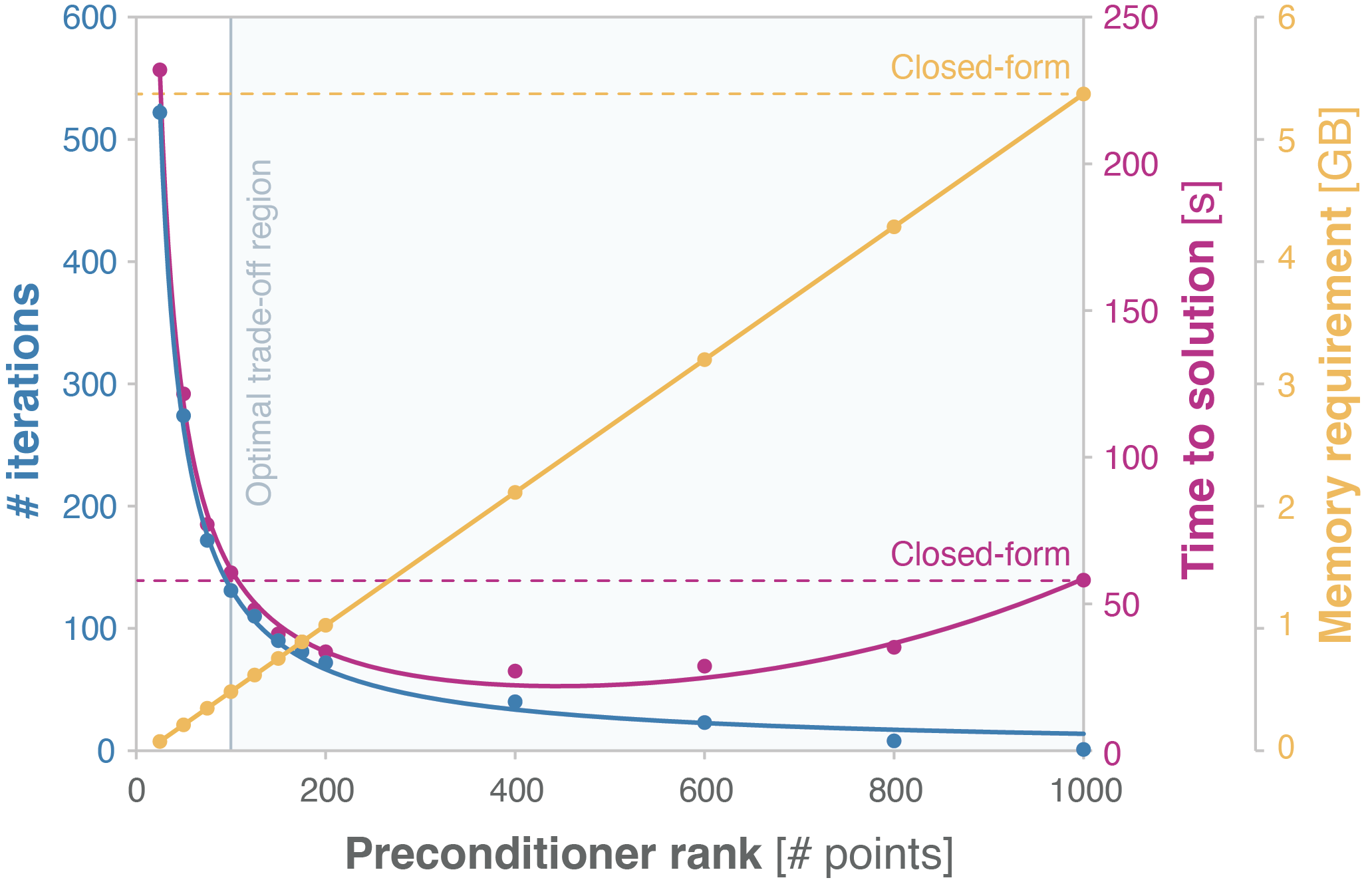}
  \caption{Number of iterations required to train a sGDML model for malonaldehyde (sampled at 500~K, 1k training points, $\sigma = 12$, target residual error below $10^{-4}$) as a function of preconditioner strength (number of inducing points). The iterative solver allows a trade-off between computational time and memory complexity. While the time to construct the preconditioner (Eq.~\ref{eq:preconditioner}) scales cubically, the increase in memory demand is linear in $k$. Using all training points for preconditioning ($\mathbf{P}^{-1} = \mathbf{K}^{-1}_\lambda$) gives the analytic solution (Cholesky decomposition) in one step. A quickly converging iterative solver can however reduce training time and memory demand at the same time (light blue region).}
 \label{fig:convergence_tests}
\end{figure}

To begin assessing the effectiveness of our approach, we investigate the role of the preconditioner configuration on the convergence behavior of the CG solver. In particular, we are interested in assessing the impact of the preconditioner strength on the overall training cost. A larger number of inducing points increases the time and memory requirements to construct $\mathbf{P}^{-1}$, but also facilitates convergence in fewer steps. The purpose of this test is to examine how effectively the critical memory restrictions of closed-form solvers can be lifted in practice. The type of chemical system represented by the training dataset is largely irrelevant for this experiment, as the convergence behavior is mainly determined by the quickly decaying spectrum of the kernel matrix~\cite{scholkopf1998nonlinear}, which is characteristic for strongly interacting many-body systems (see Fig.~\ref{fig:matrix_spectra}).

We have representatively sampled 1k points from the malonaldehyde trajectory in the MD17 dataset~\cite{chmiela2017}, which were trained with increasing preconditioning strength ranging from $2.5\%$ to $100\%$. For each configuration, the runtime and memory usage of the solver are measured (including the construction of the preconditioner). Our analysis (see Fig.~\ref{fig:convergence_tests}) shows that the CG solver converges with consistent reliability when more than around $2.5\%$ of the columns are used to construct the preconditioner. In this configuration, the memory footprint is very low, but the runtime exceeds that of the closed-form solver. However, if more than $10\%$ are used, the memory and runtime complexity are both lower with the iterative training approach. Due to the cubic scaling in the construction of $\mathbf{P}^{-1}$, the marginal benefit of increasing the preconditioner strength eventually diminishes and the runtime starts increasing again, along with the memory complexity. This means that smaller preconditioners are not only cheaper to construct and store, but also actually perform better overall (see Fig.~\ref{fig:convergence_tests}).
\\

\paragraph{Scalability}

Although our solver enables significantly larger training datasets than before (see Appendix~\ref{s:md17_large_dataset_results}), we focus on the more practical example of scaling up system size. After all, the cost generating massive reference datasets overbears any speed up that a ML model could provide. At large scales, atomic interactions become potentially more complex, as they involve a broader spectrum of length-scales. It is this scenario, in which the combined scalabilty and data efficiency of a ML model is really needed.

To put the iterative sGDML solver to the ultimate test, we have generated a new set of MD trajectories (MD22) that cover systems of up to several hundred atoms. MD22 includes examples of four major classes of biomolecules and supramolecules, ranging from a small peptide with 42 atoms, all the way up to a double-walled nanotube with 370 atoms (see Table~\ref{tab:MD22_datasets}). The trajectories were sampled at temperatures between 400~K and 500~K at a resolution of $1$ fs, with corresponding potential energy and atomic forces calculated at the PBE+MBD~\cite{perdew1996generalized, tkatchenko2012accurate} level of theory. Compared to the well-established MD17 benchmark~\cite{chmiela2017}, the standard deviations of the potential energies are significantly larger, varying between $\sim8$--$77$ kcal mol$^{-1}$ (MD17: $\sim2$--$6$ kcal mol$^{-1}$). The standard deviations of the forces are however close, between $\sim21$--$28$ kcal mol$^{-1}$ \r{A}$^{-1}$ (MD17: $\sim20$--$30$ kcal mol$^{-1}$).
We set the training dataset size for each of the systems, such that root mean squared test error for predicting atomic forces is around 1 kcal mol$^{-1}$ A$^{-1}$. For some systems like the \emph{buckyball catcher} (148 atoms) or the \emph{double-walled nanotube} (370 atoms), this error is already achieved with small training set sizes of only a few hundred points. Other systems, e.g. \emph{DHA (docosahexaenoic acid)} (56 atoms), \emph{stachyose} (87 atoms) or the \emph{Ac-Ala3-NHMe} peptide (42 atoms) require several thousands of training points for the same force prediction accuracy.
The corresponding energy mean absolute errors (MAEs) range between 0.39 kcal mol$^{-1}$ (\emph{Ac-Ala3-NHMe}) and 4.01 kcal mol$^{-1}$ (\emph{double-walled nanotube}), which is in line with our previous results on MD17~\cite{chmiela2017} when normalized per atom. The (independent) random errors made for each atomic contributions to the overall energy prediction approximately propagate as the square root of sum of squares, which causes the energy error to scale with system size~\cite{taylor1997introduction}. This scaling behavior is confirmed when comparing the energy MAE per atom, which is consistently around 0.01 kcal mol$^{-1}$ for most datasets in our study.
We observe, that the complexity of the learning task is neither correlated with the number of atoms, nor the simulation temperature of the reference trajectory. Rather, the difficulty to reconstruct a force field is determined by the complexity of the interactions within the system (see Table~\ref{tab:MD22_results}).

\begin{table*}
	\centering
	\caption{sGDML prediction performance on large-scale datasets. All (test) errors are in kcal mol$^{-1}$ (\r{A}$^{-1}$) per atom (energy) or component (forces). The training set sizes were chosen such that the root-mean-square error (RMSE) of the force prediction is around 1 kcal mol$^{-1}$ \r{A}$^{-1}$. We also provide the corresponding mean absolute errors (MAEs). $\|G\|$ denotes the cardinality of the leveraged permutation group for each respective dataset (see~\cite{chmiela2018,chmiela2019thesis,chmiela2019accurate} for details).}
	\label{tab:MD22_results}
	\begin{tabular}{lrrrrrrrrr}
		\toprule
		& \textbf{Type} & \textbf{\# atoms} & $\|G\|$ & \textbf{\# train.} & \textbf{\% data} & \multicolumn{2}{c}{\textbf{MAE}} & \multicolumn{2}{c}{\textbf{RMSE}} \\
& & & & & & Energy & Forces & Energy & Forces\\
		
		\midrule
		\multicolumn{9}{l}{\textbf{Proteins}} \\

		Ac-Ala3-NHMe &
		Tetrapeptide &
		42 &
		18 &
		6k&
		7\% &
		0.0093 & 
		0.79 &
		0.012 & 
		1.21\\
		
		\midrule
		\multicolumn{9}{l}{\textbf{Lipids}} \\
		
	    DHA (docosahexaenoic acid) &
		Fatty acid &
		56 &
		6 &
		8k &
		12\% &
		0.023 & 
		0.75 &
		0.030 & 
		1.17\\
		
		\midrule
		\multicolumn{9}{l}{\textbf{Carbohydrates}} \\
		
	    Stachyose &
		Tetrasaccharide &
		87 &
		1 &
		8k &
		29\% &
		0.046 & 
		0.68 & 
		0.052 & 
		1.07\\
		
		\midrule
        \multicolumn{9}{l}{\textbf{Nucleic acids}} \\
		
		AT-AT &
		DNA base pairs &
		60 &
		36 &
		3k &
	    15\% &
		0.012 & 
		0.69 &
		0.015 & 
		1.12 \\
		
		AT-AT-CG-CG &
		DNA base pairs &
		118 &
		96 &
		2k &
		20\% &
		0.012 & 
		0.70 &
		0.015 & 
		1.22\\

		\midrule
        \multicolumn{9}{l}{\textbf{Supramolecules}} \\

	    Buckyball catcher & 
		 &
		148 &
		48 &
		600&
		10\% &
		0.0079 & 
		0.68 &
		0.0099 & 
		1.02\\
		
		Double-walled nanotube &
		 &
		370 &
		28 &
		800&
		16\% &
		0.0108 & 
		0.52 &
		0.0135 & 
		0.97\\
		\bottomrule
	\end{tabular}
\end{table*}

\paragraph{Representation of non-local interactions}

\begin{figure}[b]
	\centering
	\includegraphics[width=1.0\columnwidth]{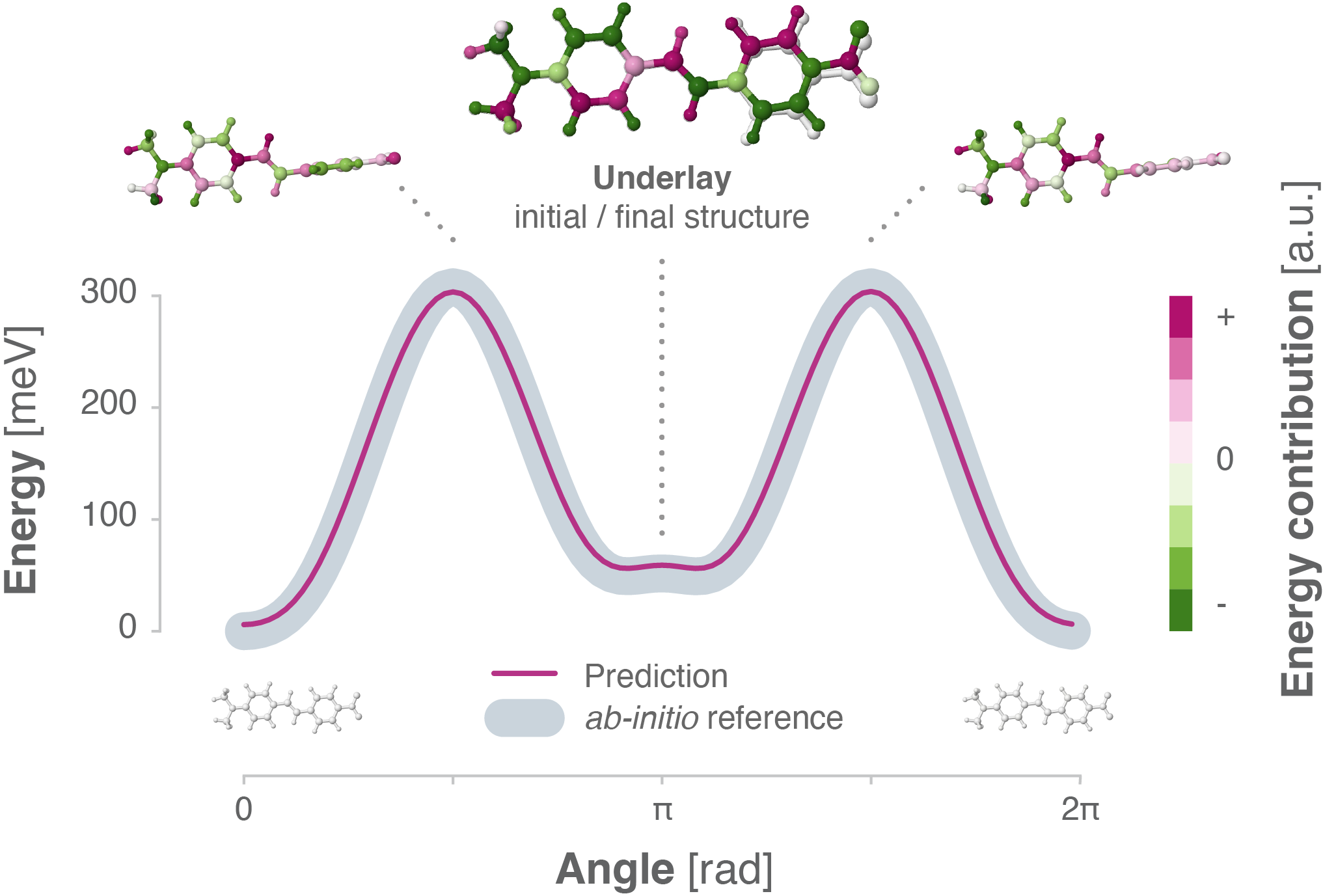}
	\caption{Energy contributions as predicted by the sGDML FF for a donor-bridge-acceptor type molecule (4-dimethylamino-4'-nitrostilbene). The energy profile for a full rotation around the single bond between the acceptor and the ethylene moiety is shown. When the conjugation of the $\pi$-system is broken upon rotating the phenyl rings by 90$^\circ$ against each other, the sGDML model predicts that the energy change is delocalized across the whole molecule.}
	\label{fig:donor_acceptor}
\end{figure}

To investigate whether the trained sGDML models provide chemically meaningful predictions, we apply sGDML to a donor-bridge-acceptor type molecule consisting of two phenyl rings connected by an \textit{E}-ethylene moiety forming a conjugated $\pi$-system (bridge) (see Section~\ref{ssec:donor_bridge-acceptor_dataset} for details on this dataset). The phenyl rings are substituted in \textit{para}-position with an electron-donating dimethylamine group (donor) and an electron-withdrawing nitro group (acceptor), respectively. When the phenyl rings are coplanar, electrons are delocalized over the whole molecule and can freely ``flow'' from donor to acceptor. However, when the two phenyl rings are rotated against each other, the conjugation of the $\pi$-orbitals is broken and the favorable interaction between donor and acceptor is lost, increasing the potential energy of the molecule. A chemically meaningful model should predict that this energy change is delocalized over the whole $\pi$-system (as opposed to explaining it by local changes in the vicinity of the center of the rotation). To get a qualitative understanding of how these interactions are handled within the sGDML model, we investigate how individual atoms contribute towards the prediction. 
Being a linear combination of pairwise correlations between atoms, a partial evaluation of the model reveals the atomic contributions to each prediction (see Fig.~\ref{fig:donor_acceptor}).
We observe that all atoms in the system participate in generating the prediction with sGDML, which would not be possible with a model that partitions the energy into localized atomic contributions. Figure~\ref{fig:donor_acceptor} demonstrates that a sGDML model learns to delocalize changes in energy upon ring rotation across the whole molecule, which is in accordance with chemical intuition. Note that, starting from the global minimum structure of this system, rotating by $\pi$ does not return to the starting position (despite the apparent symmetry of the molecule), because of a slight asymmetry about the central C=C bond (see overlay of structures in Figure~\ref{fig:donor_acceptor}). Thus, a full rotation is necessary to return to the starting point, explaining the somewhat counter-intuitive rotational energy profile.

\section{Molecular dynamics}

One of the biggest advantages of employing MLFFs is that they can enable accurate large-scale simulations. Here, we test our sGDML FFs by running nanosecond-long classical MD and path integral MD (PIMD) simulations for the \emph{double-walled carbon nanotube} saturated with hydrogen atoms at its edges. All simulations were run at a constant temperature of 300~K with a Langevin thermostat and a time-step of 0.2~fs. The number of beads of the PIMD simulations was set to 16. 

To confirm the reliability of any MLFF, it is first and foremost essential to assess its capability to yield stable (PI)MD simulations. In this regard, Fig.~\ref{fig:cumulative_E} shows the cumulative potential-energy ($V_{step}=\frac{1}{N_{step}}\sum_{k=1}^{N_{step}}V_{k}$; where $N_{step}$ is the number of completed time steps, and $V_{k}$ is the potential-energy at the $k^{\text{th}}$ step) along both MD and PIMD simulations. After thermalization (roughly 500~ps for MDs and 100~ps for PIMDs), the simulations reach equilibrium.
\begin{figure}[hb!]
	\centering
	\includegraphics[width=1.0\columnwidth]{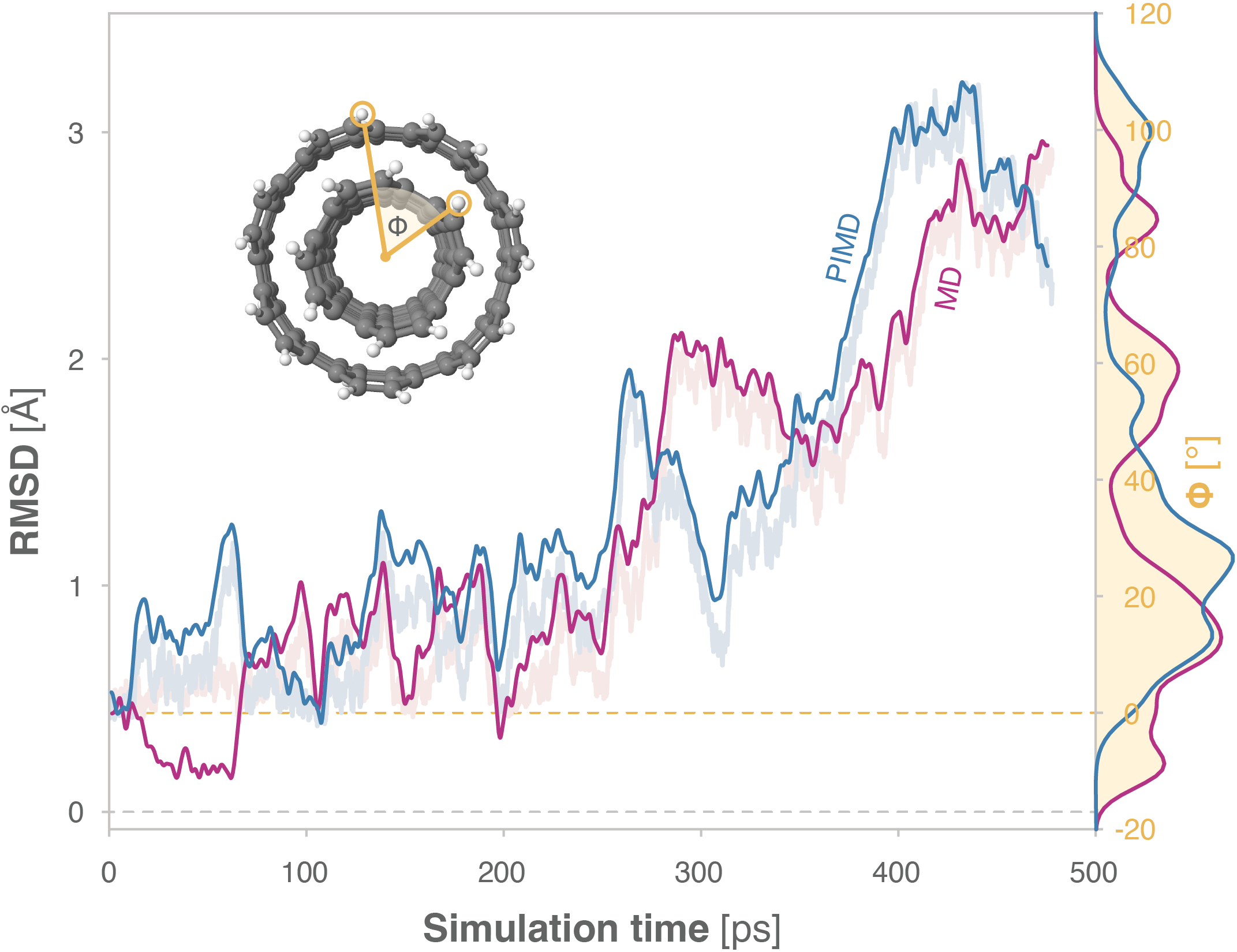}
	\caption{Instantaneous root-mean squared deviation (RMSD in \r{A}ngstr\"om; transparent lines in the background) and angle of the relative rotation ($\Phi$ in degrees) between both nanotubes as a function of simulation time (in ps). The RMSDs were computed with respect to the initial configurations used for the simulations.
	}
	\label{fig:nano_RMSD_Rotation}
\end{figure}
Next, we compare the difference in geometry fluctuations between MD and PIMD simulations in Fig.~\ref{fig:nano_RMSD_Rotation}, measured by root-mean-squared deviations (RMSDs) from the initial geometry.
The RMSDs of the PIMD present a series of peaks that are higher than those observed in the classical MD simulation. The first of such peaks appears at $\sim$60~ps and then there is one every 100~ps (the highest one corresponding to a RMSD greater than 3.0~\AA). The origin of these RMSD fluctuations is the relative angle of rotation ($\Phi$) of the inner nanotube with respect to the outer one (see Fig.~\ref{fig:nano_RMSD_Rotation}). 
The outer and inner nanotubes have a 7- and 4-fold axis of rotation (the axis parallel to the nanotubes), respectively, meaning that we have the same configuration every $\sim$13$^{\circ}$. However, RMSDs are dependent on atom indices and uncoupled rotations of the nanotubes lead to a different arrangement of the atoms with respect to each other. This causes the increments observed in RMSDs along the simulations since no other degrees of freedom fluctuate as much as the angle $\Phi$.
Indeed, we observe a strong correlation between the RMSD variations and the evolution of the angle $\Phi$ in a scale of 360$^{\circ}$ (a full rotation).
The differences in the values of $\Phi$ between the MD and PIMD simulations suggest that a coupling of nuclear quantum effects (NQEs) and long-range interactions (resembling existing studies on the stability of different aspirin crystal polymorphs~\cite{reilly2014role}) eases the rotation of one of the nanotubes with respect to the other. The distribution of values of $\Phi$ further confirms that NQEs smoothen the rotational profile of the nanotubes. 
While in the PIMD values of $\Phi$ from 40$^{\circ}$ to 80$^{\circ}$ are equally sampled, in the MD one can observe two pronounced peaks at around 60$^{\circ}$ and 80$^{\circ}$.
It is important to note that the rather large time-scale (100~ps) between each of these rotations indicates that this motion corresponds to low-frequency vibrational modes.

As a final demonstration of the capability of sGDML MLFF models for providing insights into large systems, we computed the molecular vibrational spectra of the \emph{buckyball catcher} and the \emph{double-walled nanotube} from both MD and PIMD simulations (see Fig.~\ref{fig:spectra}). These spectra correspond to velocity auto-correlation functions.
The spectra feature peaks corresponding to \mbox{=C-H} stretching (at around 3000~cm$^{-1}$) and bending (close to 1000~cm$^{-1}$) modes, as well as those that correlate to \mbox{C=C} vibrations at around 500 and 1500~cm$^{-1}$ accounting for expansions and contractions of the buckyball, the ``hands'' of the catcher and the nanotubes (for instance, see Refs.~\cite{Hare1991IR,Choi2000Vibrational} for a discussion of the vibrational spectra of the buckyball).

Although MD and PIMD simulations provide similar spectra, the inclusion of NQEs yields more accurate frequencies. Namely, nuclear quantum delocalization leads to a shift of the \mbox{=C-H} stretching mode, which puts the peak closer to 3000~cm$^{-1}$. This value is in agreement to that of other aromatic and $\pi$-$\pi$ interacting systems. For instance, with experimental and theoretical values of the fundamental C-H stretching modes of benzene and the benzene dimer (mainly the $\nu_{13} (\mathbf{B}_{1u})$ mode)~\cite{erlekam2006experimental}. Hence, PIMD simulations capture some of the anharmonic behavior of the systems and correct the overestimated value of $\approx$~3100~cm$^{-1}$ in the classical MD.
Differently to \mbox{=C-H} stretching, the parts of the spectra corresponding to ``long-range'' vibrations (i.e., low-frequency modes) are, in general, consistent among MD and PIMD. This agreement aligns with the fact that both the \emph{buckyball catcher} and the \emph{double-walled nanotube} are relatively symmetric systems and that the differences between configuration spaces sampled by MD and PIMD simulations are mostly of local nature. 
Therefore, even though NQEs promote a low-frequency mode, such as the relative rotation between the nanotubes, these modes are smoothed out in the low-frequency part of the vibrational spectrum.

\section{Conclusion}

Kernel-based FFs are known to be sample efficient, but believed to be limited in their ability to scale well with training set or system size. A key reason is the common use of direct solvers, which factorize the kernel matrix in order to solve the associated optimization problem to train the model. While this approach is numerically stable, it quickly incurs a prohibitive memory and runtime complexity. This dilemma can be evaded using iterative solvers, which essentially allow kernel-based models to be trained similar to neural networks. However, the straightforward application of iterative solvers is difficult when large kernel matrices are involved, due to their notoriously poor numerical conditioning.

In this work, we propose an iterative scheme that enables the robust application of sGDML to significantly larger systems (both, in terms of training set and system size) without introducing any approximations to the original model.
This is achieved with a numerical preconditioning scheme which drastically reduces the conditioning number of the learning problem, enabling rapid convergence of a CG iteration.
With this advance, we are now able to apply our kernel-based model to large-scale learning tasks that have previously only been accessible to neural networks, while carrying over the sample efficiency and accuracy of sGDML. We attribute the latter to the model's unique ability to represent global interactions on equal footing with local interactions, as a series of numerical experiments demonstrate. Now, molecular systems that exhibit phenomena with far-reaching characteristic correlation lengths can be studied in long-timescale MD simulations.

Breakthroughs in MLFF development are often driven by the creation of benchmark datasets that offer ever evolving challenges. Early quantum chemistry datasets such as QM7~\cite{rupp2012fast}, MD17~\cite{chmiela2017}, the valence electron densities for small organic molecules in Ref.~\cite{brockherde2017bypassing}, ISO17~\cite{schutt2017}, SN2~\cite{unke2019}, or SchNOrb~\cite{schutt2019unifying, gastegger2020deep}, focused on defining useful inference problems and opportunities in quantum chemistry, whereas later benchmarks (e.g. QM7-X~\cite{hoja2021qm7}) steered the field towards developing more robust transferable models. The MD22 benchmark dataset developed in this work now offers new challenges for atomistic models with regard to molecular size and flexiblity that could further advance research on novel MLFF architectures similarly to previous datasets of quantum-mechanical calculations.

Our technical development allows future cross-fertilization between both MLFF development approaches: Now, kernel-based MLFFs can capitalize on the massive parallelism available on GPUs and the software infrastructure that enabled scalability of deep neural networks. On the other hand, modeling principles from kernel methods inspire the development of new architectures such as transformers using self-attention~\cite{frank2021detect, frank2022so3krates} and pave the way out of overly restrictive localization assumptions. The exclusive use of on-the-fly model evaluations by iterative solvers also represents a paradigm shift in the way kernel-based MLFFs are typically trained, which opens up new avenues for further developments. We have recently shown, how this makes them amenable to strong differential equation constraints via algorithmic differentiation techniques to simplify descriptor development and further improve data efficiency~\cite{schmitz2022}. With the ability to reconstruct MLFFs for larger systems, the need for better management of the growing set of molecular features arises. To this end, we have recently proposed a novel descriptor pruning scheme to contract trained models and make them easier to evaluate~\cite{kabylda2022towards}. One could also envision a systematic construction of local and non-local fragments (by generalizing from non-interacting to \emph{interacting} amons~\cite{huang2020quantum}) that would enhance the scalability and transferability of global MLFFs. Future research will furthermore explore our significantly better scaling behavior across a broad range of application fields in the physical sciences.

{\bf Funding statement}: SC and KRM acknowledge support by the German Federal Ministry of Education and Research (BMBF) for BIFOLD (01IS18037A). KRM was partly supported by the Institute of Information \& Communications Technology Planning \& Evaluation (IITP) grants funded by the Korea government(MSIT) (No. 2019-0-00079, Artificial Intelligence Graduate School Program, Korea University and No. 2022-0-00984, Development of Artificial Intelligence Technology for Personalized Plug-and-Play Explanation and Verification of Explanation) and by the German Federal Ministry for Education and Research (BMBF) under Grants 01IS14013B-E, 01GQ1115. AT, VVG and AK acknowledge support from the European Research Council (ERC) Consolidator Grant BeStMo and the Luxembourg National Research Fund (FNR) (Grant C19/MS/13718694/QML-FLEX and AFR PhD Grant 15720828). HES acknowledges support from DGTIC-UNAM under Project LANCAD-UNAM-DGTIC-419. Furthermore, we thank IPAM for their warm hospitality and inspiration while finishing this manuscript.



{\bf Data, model and software availability}:
All datasets, pre-trained FF models and software used in this work are available at \texttt{www.sgdml.org}.

\bibliography{main}

\onecolumngrid
\clearpage

\appendix

\setcounter{equation}{0}
\setcounter{figure}{0}
\setcounter{table}{0}
\setcounter{page}{1}
\makeatletter
\renewcommand{\theequation}{S\arabic{equation}}
\renewcommand{\thefigure}{S\arabic{figure}}
\renewcommand{\thetable}{S\Roman{table}}

\newpage
\widetext
\begin{center}
\textbf{\large Supplementary Information}
\end{center}

\section{MD22 dataset properties}


\begin{table}[H]
	\centering
	\caption{Computational details of the MD22 datasets. All calculation were performed using the FHI-aims~\cite{blum2009ab} electronic structure software, in combination with i-PI~\cite{kapil2019pi} for the MD simulations. The potential energy and atomic force labels were calculated at the PBE+MBD~\cite{perdew1996generalized, tkatchenko2012accurate} level of theory. The keywords \emph{light} and \emph{tight} denote different basis set options in FHI-aims. All trajectories were sampled at a resolution of $1$ fs. \emph{Coeff.} refers to the friction coefficient used for the global Langevin thermostat (in fs) or the effective mass for the Nosé-Hoover thermostat (in cm$^{-1}$), respectively.
	}
	\label{tab:MD22_datasets_settings}
	\begin{tabular}{lrrrrr}
		\toprule
		\textbf{Dataset} & \textbf{Temp. [K]} & \textbf{ Level of theory } &  \textbf{Thermostat} &  \textbf{Coeff.} \\
		\midrule
		
		Ac-Ala3-NHMe &
		500&
		PBE+MBD / tight&
		Global Langevin &
		2 \\

		DHA (Docosahexaenoic acid) &
		500&
		PBE+MBD / tight&
		Nosé-Hoover &
		1700 \\
		
	    Stachyose &
		500&
		PBE+MBD / tight&
		Nosé-Hoover&
		1700 \\
		
		DNA base pair (AT-AT) &
		500&
		PBE+MBD / tight&
		Global Langevin&
		2 \\

		DNA base pair (AT-AT-CG-CG) &
		500&
		PBE+MBD / tight&
		Global Langevin&
		2 \\

		Buckyball catcher&
		400&
		PBE+MBD / light&
		Nosé-Hoover&
		1700 \\
		
		Double-walled nanotube &
		400&
		PBE+MBD / light&
		Nosé-Hoover&
		1700 \\
		\bottomrule
	\end{tabular}
\end{table}

\begin{table}[H]
	\centering
	\caption{Properties of the MD22 datasets. 
	Energies are in kcal mol$^{-1}$ A$^{-1}$, forces in kcal mol$^{-1}$ A$^{-1}$.}
	\label{tab:MD22_datasets}
	\begin{tabular}{lrrrrrr}
		\toprule
		\textbf{
		Dataset} & \textbf{Formula} & \textbf{Size}  & \multicolumn{2}{l}{\textbf{Energies}} & \multicolumn{2}{l}{\textbf{Forces}} \\ 
        & & & Range & Variance & Range & Variance \\
		\midrule
		
		Ac-Ala3-NHMe &
		C$_{12}$H$_{22}$N$_{4}$O$_{4}$&
        85,109 &
		102.19 &
		67.30 &
		437.99 &
		678.02\\

		DHA (Docosahexaenoic acid) &
		C$_{22}$H$_{32}$O$_{2}$&
		69,753 &
		75.71 &
        91.31 &
		420.07 &
		673.99\\
		
	    Stachyose &
		C$_{24}$H$_{42}$O$_{21}$&
		27,272 &
		106.15 &
        189.33 &
		426.08 &
		655.55	\\
		
		DNA base pair (AT-AT) &
		C$_{20}$H$_{22}$N$_{14}$O$_{4}$ &
		20,001 &
		139.29 &
        120.09 &
		444.01 &
		779.81\\

		DNA base pair (AT-AT-CG-CG) &
		C$_{38}$H$_{42}$N$_{30}$O$_{8}$&
		10,153 &
		243.50 &
        246.64 &
		407.55 &
		768.13\\

		Buckyball catcher&
		C$_{120}$H$_{28}$&
		6,102 &
		378.08 &
        576.85 &
		287.77 &
		450.12\\
		
		Double-walled nanotube &
		C$_{326}$H$_{44}$&
		5,032 &
		674.50 &
        5950.68 &
		315.39 &
		546.30\\
		\bottomrule
	\end{tabular}
\end{table}

\section{Donor-bridge-acceptor dataset}
\label{ssec:donor_bridge-acceptor_dataset}

The reference data for the donor-bridge-acceptor example was generated by normal mode sampling~\cite{smith2017ani} at 300~K with random rotations applied to individual bonds. Energies and forces were calculated using the semi-empirical GFN2-xTB method~\cite{bannwarth2019gfn2}. The energy profile in Fig.~\ref{fig:donor_acceptor} shows a rotation around a single bond, with all other bond distances and angles fixed at their equilibrium values.

\section{Scaling to larger training dataset sizes}
\label{s:md17_large_dataset_results}

We have also investigated the scaling behavior of sGDML to larger training datasets, in addition to larger system sizes. Until now, kernel-based MLFFs only scaled to around 1k training points on the MD17 dataset, while a training set size of 50k has emerged as popular choice for models based on neural networks~\cite{unke2020}. Most neural network architectures are not fully converged at 1k, requiring more training data to achieve competitive generalization errors~\cite{christensen2020fchl}, although more data efficient neural-network-based FFs are starting to appear~\cite{batzner2021se, schutt2021equivariant, frank2022so3krates}. With our proposed numerical training approach, large-scale problems now become accessible to kernel-based models, making a direct comparison of kernel-based and neural-network-based potentials possible on a wide range of training dataset sizes.

For this purpose, we have trained (s)GDML models for all molecules in MD17 with dataset sizes of 1k and 50k using a preconditioner based on 75 inducing points in each case. For the largest molecule (aspirin) at 50k, the iterative solver only uses $0.15$ per mil of the memory compared to a closed-form solver, allowing an increase of the learning problem size by two orders of magnitude on the same hardware. All models have been converged to a training error of $10^{-4}$ kcal mol$^{-1}$ A$^{-1}$, which was reached between $\sim800$ (benzene) and $\sim2700$ (ethanol) iteration steps. The test results are shown in Table~\ref{tab:md17_results}.

We remark that such massive training set sizes are infeasible for any practical FF reconstruction task, as the computational cost of generating the reference dataset overbears any speed up that the ML model can provide. Rather, the key performance metric of any ML potential is its training data efficiency, i.e. how quickly a certain test performance can be reached. The only purpose of this exercise is to demonstrate the stability of our iterative solver for large-scale problems.

\begin{table}[H]
	\centering
	\caption{sGDML prediction performance on the extended MD17 datasets~\cite{chmiela2017, chmiela2018} for 50k training points. All mean absolute errors (MAE) test errors are in kcal mol$^{-1}$ (\r{A}$^{-1}$). Energy errors are given for the complete structure, whereas force errors are per component.}
	\label{tab:md17_results}
	\begin{tabular}{lrrrlrr}
		\toprule
		\textbf{Dataset} & \textbf{\# atoms} & \multicolumn{2}{l}{\textbf{MAE}}  \\
 & & Energy & Forces  \\
		
		\midrule
		\multicolumn{4}{l}{\textbf{50k}}\\
		\midrule
		
        Benzene\textsubscript{2017} &
		12 &
		0.069 &
		0.140\\
		
	    Uracil &
		12 &
		0.103 &
		0.027\\
		
		Naphthalene &
		18 &
		0.114 &
		0.029\\
		
		Aspirin &
		21 &
		0.122 &
		0.047\\
		
		Salicylic acid &
		16 &
		0.105 &
		0.039\\
		
		Malonaldehyde &
		9 &
		0.074 &
		0.083\\
		
		Ethanol &
		9 &
		0.050 &
		0.058 \\
		
		Toluene &
		15 &
		0.092 &
		0.033 \\
		
		\midrule
		
		Paracetamol &
		20 &
		0.113 &
		0.051 \\
		
		Azobenzene &
		24 &
		0.137 &
		0.066 \\
		
		\bottomrule
	\end{tabular}
\end{table}

\section{Parametric complexity of the models}
\label{s:parametric_complexity}

In contrast to neural network architectures, sGDML FFs are based on non-parametric GP models. The parameter set of non-parametric models is not fixed, but adapts to the complexity and amount of available training data. This evokes concerns about the evaluation speed of large-scale sGDML models. In practice, the parametric complexity of sGDML FFs is however around one order of magnitude lower than that of neural-network-based FFs with comparable accuracy.

Table 1 in Ref.~\cite{frank2022so3krates} shows, that the parameter set sizes range from 500k (NewtonNet) to 3M (SpookyNet, NequIP) for neural-network-based FFs trained on the MD17 datasets using 1k training points. Table 2 in Ref.~\cite{fu2022forces} shows parameter set sizes ranging from 120k (SchNet), ~1M (NequIP) to ~12M (ForceNet) for 10k training points on the same datasets. In comparison, the largest sGDML model (for MD17-aspirin, 21 atoms) only uses 63k parameters for 1k training points and 630k parameters for 10k training points. Moreover, sGDML models are linear in their parameters and therefore trivial to parallelize. In contrast, deep neural network architectures have a hierarchical structure, which is computationally more expensive to evaluate, even when the number of parameters is similar.

\section{Cholesky decomposition}
\label{s::cholesky}

Due to the symmetric positive-definite properties of the kernel function, the system $\mathbf{\alpha} = (\mathbf{K} + \lambda \mathbb{I})^{-1}\mathbf{y}$ can be solved using a Cholesky decomposition into a product of lower triangular factors $\mathbf{L} \in \mathbb{R}^{m \times m}$~\cite{fine2001efficient, bach2002kernel},
\begin{equation}
\mathbf{\alpha} = (\mathbf{K} + \lambda \mathbb{I})^{-1}\mathbf{y} = (\mathbf{L}\mathbf{L}^{\top})^{-1}\mathbf{y},
\label{eq:cholesky}
\end{equation}
and subsequent forward and backward substitution.

\section{Approximate $\lambda$-ridge leverage scores}
\label{s::app_lev_scores}


We use the Nystr\"om method to project the (symmetric and positive semidefinite) kernel matrix $\mathbf{K}$ onto a subset of its columns, to obtain a memory efficient and easily invertible approximation of $\mathbf{K}_\lambda$ as preconditioner for the learning problem. The choice of inducing columns determines the quality of the approximation, i.e.\ how well the eigenvalue spectrum of the original kernel matrix is represented.
Here, we use an extended notion of leverage score that is tailored to the context of GPs to determine a good set of inducing points. These so called \emph{statistical leverage scores} depend on a regularization parameter $\lambda$ that 
diminishes the importance of small eigendirections~\cite{alaoui2015fast, rudi2018fast}:
\begin{equation}
\tau_{i}(\lambda)=[\mathbf{K}(\mathbf{K} + \lambda \mathbb{I})^{-1}]_{ii}.
\end{equation}
Note, that $\tau_i(\lambda)$ is the $i$-th diagonal entry of the matrix product. The exact computation of $\tau_i(\lambda)$ is however not practical as it requires inverting $\mathbf{K} + \lambda \mathbb{I}$, which is as expensive as solving the original learning problem~\cite{cohen2015uniform}. Instead, approximate \emph{$\lambda$-ridge leverage scores} are obtained based on another Nystr\"om approximation using simple uniform column sampling. With $\mathbf{B} \mathbf{B}^{\top} = \mathbf{K}_{mk} \mathbf{K}_{kk}^{-1} \mathbf{K}_{mk}^{\top}$, we have 
\begin{equation}
\tilde{\tau}_{i} (\lambda)=\mathbf{b}_{i}^{\top}\left(\mathbf{B}^{\top} \mathbf{B} + \lambda \mathbb{I}\right)^{-1} \mathbf{b}_{i},
\end{equation}
as approximate measure for the importance of each column $i$. Here, $\mathbf{b}_i$ is the $i$-th row of the Cholesky factor $\mathbf{B}$. The inducing columns for our preconditioner are then sampled according to this leverage score distribution. It has been shown that this construction approximates $\mathbf{K}_\lambda$ with constant probability within a small relative error in comparison to the best rank-$k$ approximation via eigenvalue decomposition~\cite{drineas2008relative, kumar2009sampling}, but at significantly lower computational cost.

\section{Implementation details}
\label{s::implementation_details}

The performance and effectiveness of our solver crucially depends on a careful implementation, as it involves operations that are highly susceptible to numerical error. Here, we give an overview of the most important practical considerations in our reference implementation.
\\
\paragraph{Memory requirement}

At every iteration, the preconditioned CG algorithm computes two matrix products: $\mathbf{K}_\lambda \mathbf{\alpha}_t$ (see Eq.~\ref{eq:normal_eq_gradient}) and another one involving $\mathbf{P}^{-1}$. To avoid storing the $m \times m$ kernel matrix $\mathbf{K}_\lambda$, we use the NumPy \texttt{LinearOperator} interface to perform matrix-vector multiplications on-the-fly. This reduces the memory complexity of $\mathbf{K} \mathbf{\alpha}_t$ to $\mathcal{O}(m)$ as only the resulting vector needs to be stored. $\mathbf{P}^{-1}$ is relatively expensive to construct and therefore retained in memory, albeit in factorized form at a smaller memory cost of only $\mathcal{O}(mk)$.
\\
\paragraph{Preconditioner construction} A straightforward evaluation of the Woodbury matrix identity in Eq.~\ref{eq:preconditioner} is known to be numerically unstable~\cite{foster2009stable}. To construct $\mathbf{P}^{-1}$, we have to take a computational detour. First, the Cholesky decomposition of the symmetric positive definite sub-matrix $\mathbf{K}_{kk} = \mathbf{L}_{kk}\mathbf{L}_{kk}^{\top}$ (Eq.~\ref{eq:nystrom_approx}) is generated to define $\mathbf{\bar{K}}_{mk} = \mathbf{K}_{mk}  \mathbf{L}_{kk}^{-\top}$~\cite{wahba1990spline}. Using another Cholesky step $\lambda\mathbb{I}+\mathbf{\bar{K}}^{\top}_{mk} \mathbf{\bar{K}}_{mk} = \mathbf{D}_{kk} \mathbf{D}_{kk}^{\top}$, we then construct
\begin{equation}
\mathbf{P}^{-1} = \lambda^{-1} \left[\mathbb{I} - \mathbf{\bar{K}}_{mk}\mathbf{D}_{kk}^{-\top} \mathbf{D}_{kk}^{-1}\mathbf{\bar{K}}_{mk}^{\top}\right].
\end{equation}

However, evaluating the product $\mathbf{\bar{K}}^{\top}_{mk} \mathbf{\bar{K}}_{mk}$ is not advisable, because it squares the condition number of the term, which can introduce rounding errors that diminish the effectiveness of the preconditioner. We therefore perform a numerically more robust indirect Cholesky decomposition of the inverse in Eq.~\ref{eq:preconditioner} via thin QR factorization~\cite{golub2013matrix, wahba1990spline, seeger2003fast, foster2009stable, eriksson2018scaling}:
\begin{equation}
\left[\begin{array}{c}
\mathbf{\bar{K}}_{mk} \\
\sqrt{\lambda}\mathbb{I}
\end{array}\right]
= \mathbf{Q}\left[\begin{array}{c}
\mathbf{D}_{kk}^{\top} \\
\mathbf{0}
\end{array}\right]
\end{equation}
This decomposition can be computed efficiently without storing $\mathbf{Q}$, albeit at a slightly higher cost than a direct Cholesky decomposition. Throughout these calculations, the memory complexity remains constant at $\mathcal{O}(mk)$. $\mathbf{P}^{-1}$ is then applied to vectors via the NumPy \texttt{LinearOperator} interface without expanding the product.
\\
\paragraph{Initial guess $\mathbf{\alpha}_{0}$} We use $\mathbf{\alpha}_{0} = \mathbf{0}$ as initial guess, because a non-zero initialisation can adversely affect convergence of Krylov subspace methods (see Ref.~\cite{liesen2013krylov}, chapter 5.8.15).
\\
\paragraph{CG solver restarts} With perfect arithmetic, the CG algorithm guarantees monotonically improving approximations of $\mathbf{\alpha}_{t}$. However, the series of optimization steps may eventually become non-orthogonal after many iterations, due to the accumulation of rounding errors. This can slow down or even completely stall convergence in practice~\cite{powell1977restart}. We monitor the average progress of the solver towards minimizing the residual within a rolling time window and break the cycle by restarting the CG algorithm using the latest $\mathbf{\alpha}_{t}$ as initial guess, if necessary. Slow progress towards the solution can also be a consequence of insufficient preconditioner performance. Our implementation dynamically increases the number of inducing columns to adjust the strength of the preconditioner before every restart.

\begin{figure}[ht!]
	\centering
	\includegraphics[width=0.5\columnwidth]{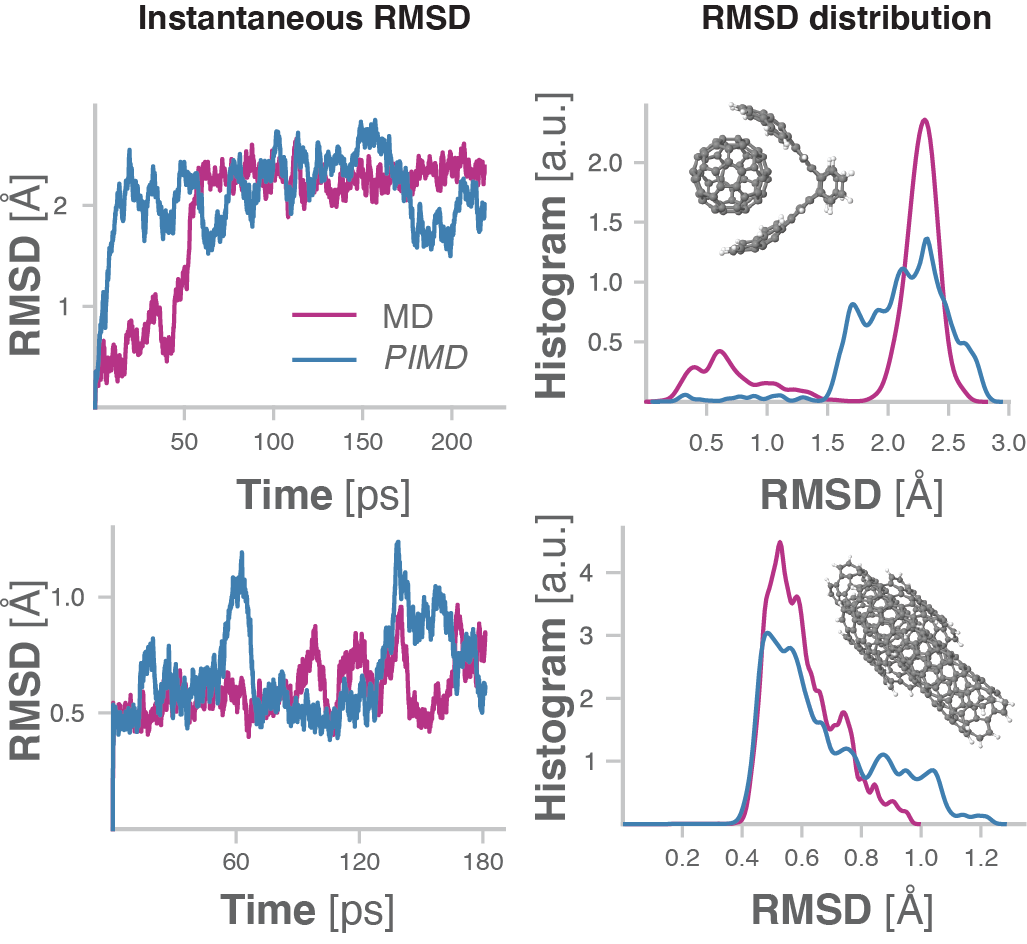}
	\caption{Instantaneous root-mean squared deviation (RMSD; in \r{A}ngstr\"om) as a function of simulation time (in ps) and distributions of RMSDs for the MD and PIMD simulations of the buckyball catcher and the double-walled nanotube.}
	\label{fig:RMSDs}
\end{figure}

\begin{figure}[ht!]
	\centering
	\includegraphics[width=0.5\columnwidth]{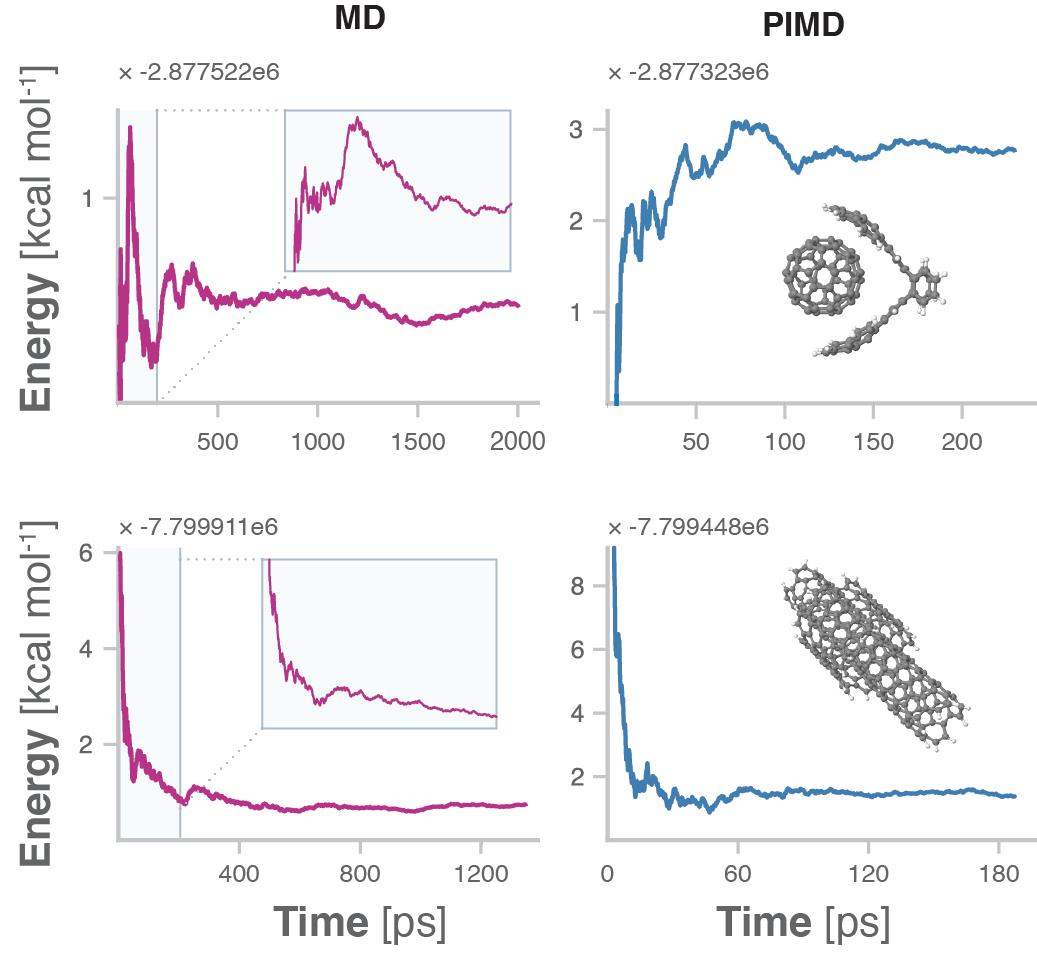}
	\caption{The cumulative potential-energy (in kcal mol$^{-1}$) as a function of simulation time (in ps) for the buckyball catcher and the double-walled nanotube, along classical MD and PIMD simulations obtained with sGDML FFs. The simulations were carried out until the cumulative energy remained approximately constant. The classical MD plots contain zoomed-in sections spanning a simulation time equivalent to the length of the PIMD trajectories.}
	\label{fig:cumulative_E}
\end{figure}

\begin{figure}[ht!]
	\centering
	\includegraphics[width=0.5\columnwidth]{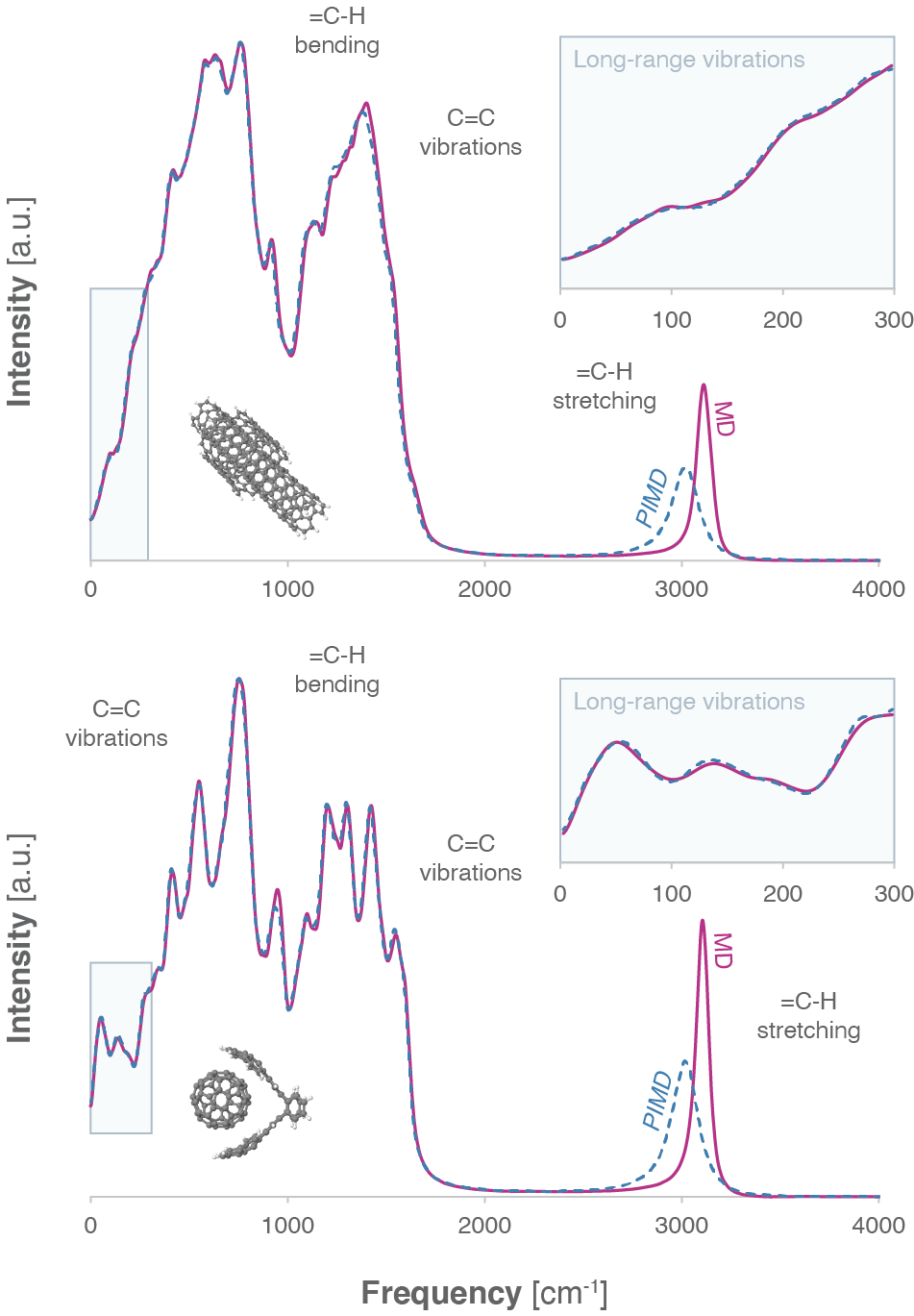}
	\caption{Molecular spectra of the double-walled nanotube and the buckyball catcher, as obtained from the velocity autocorrelation function for MD and PIMD simulations. For the sake of consistency, the MD trajectories have been cropped to the same length for this plot. The final length was determined by the total number of steps in the PIMD simulations and does not affect their convergence.}
	\label{fig:spectra}
\end{figure}

\end{document}